\documentclass[final,5p,times,twocolumn,number]{elsarticle}

\usepackage{scrextend}
\usepackage{epsfig}
\usepackage{amsmath}
\usepackage{mathtools}
\usepackage{amsfonts}
\usepackage{amssymb}
\usepackage{color}
\usepackage{graphicx}
\usepackage{bm}
\usepackage{ulem}

\definecolor{navyblue}{rgb}{0.0, 0.0, 0.5}
\definecolor{royalblue}{rgb}{0.25, 0.41, 0.88}
\definecolor{cadmiumgreen}{rgb}{0.0, 0.42, 0.24}
\definecolor{blue-violet}{rgb}{0.54, 0.17, 0.89}
\definecolor{darkviolet}{rgb}{0.58, 0.0, 0.83}
\definecolor{orange(colorwheel)}{rgb}{1.0, 0.5, 0.0}

\usepackage{hyperref}
\hypersetup{
    colorlinks=true, 
    linkcolor=royalblue, 
    citecolor=magenta}

\usepackage{booktabs}
\usepackage{multirow}
\usepackage{dcolumn}
\usepackage{colortbl}

\usepackage{stackengine}

\biboptions{sort&compress}

\journal{Physics Letters B}

\begin{document}

\title{Signatures of Modified Gravity on Linear Scales in a Dynamical Dark Energy Background}

\author{Yo Toda}
\ead{toda.yo@kochi-tech.ac.jp}
\address{Department of Data \& Innovation, Kochi University of Technology, Tosayamada 782-8502, Japan \looseness=-1}

\author{Adrià Gómez-Valent}
\ead{agomezvalent@icc.ub.edu}
\address{Departament de Física Quàntica i Astrofísica (FQA) and Institut de Ciències del Cosmos (ICCUB), Universitat de Barcelona (UB), c. Martí i Franqués, 1, 08028 Barcelona, Catalonia, Spain}

\begin{abstract}
Cosmological data from the cosmic microwave background (CMB), baryon acoustic oscillations, and Type Ia supernovae suggest that the component driving the accelerated expansion of the Universe may be dynamical at the $\sim 2.5$-$3\sigma$ CL. The best-fit CPL model produces a level of cosmic structure similar to that of $\Lambda$CDM, with both models exhibiting mild tension with redshift-space distortion data. In this {\it Letter}, we parametrize possible departures of the effective gravitational coupling from Newton's constant in the late Universe, below a comoving scale $\lambda_c$, using two redshift bins, $0 \leq z < 1$ and $1 \leq z \leq 3$. We then determine the optimal values of $\lambda_c$ and the amplitude of these deviations from General Relativity, assuming a background with dynamical dark energy in CPL form. We find that, in order to achieve the required suppression of structure growth at low redshifts while remaining consistent with CMB constraints -- primarily from the late-time ISW effect at low $\ell$ and lensing at high $\ell$ -- we must require modified gravity effects to have a characteristic scale of $\lambda_c \lesssim 150\,\mathrm{Mpc}$ (95\% CL). This places the relevant scales squarely within the range probed by galaxy surveys. The best-fit value is $\lambda_c \simeq 40\,\mathrm{Mpc}$, well below the 95\% CL upper limit. Using Planck PR4, DESI DR2, Pantheon+ (or DES-Dovekie) and redshift-space distortions data we confirm that a CPL background with standard gravity is moderately preferred over $\Lambda$CDM; this preference strengthens to a mildly strong level when modified gravity effects are included. This enhancement leaves the CPL parameters largely unchanged, but shifts them slightly further into the quintom region.
\end{abstract}

\maketitle

\section{Introduction}\label{sec:introduction}

The $\Lambda$CDM model has proven highly effective in explaining a wide range of cosmological phenomena across cosmic history, from inflation to the current accelerated phase (see, e.g., \cite{Huterer:2017buf}). This is despite the fact that its theoretical status remains far from complete, with several key aspects still not fully understood, including the fundamental nature of the cosmological constant $\Lambda$ \cite{Weinberg:1988cp,Sahni:1999gb,Carroll:2000fy,Peebles:2002gy,Padmanabhan:2002ji,Sola:2013gha,SolaPeracaula:2022hpd,SolaPeracaula:2026pgi} and of dark matter \cite{Bozorgnia:2024pwk}, which together constitute two of its main ingredients. The advent of precision cosmology has also revealed a number of tensions between the model and observations \cite{Perivolaropoulos:2021jda,CosmoVerseNetwork:2025alb}, challenging its phenomenological viability and possibly hinting at the need for new physics beyond it. Both theoretical considerations and observational evidence leave ample room for departures from the standard paradigm.

In fact, baryon acoustic oscillation (BAO) measurements from the Dark Energy Spectroscopic Instrument (DESI) \cite{DESI:2025zgx}, together with various Type Ia supernova (SNIa) samples \cite{Scolnic:2021amr,Rubin:2023jdq,DES:2025sig} and cosmic microwave background (CMB) data from the Planck satellite \cite{Planck:2018vyg,Efstathiou:2019mdh,Rosenberg:2022sdy}, provide hints of dynamical dark energy (DE) in the late Universe. In particular, within the assumption of standard pre-recombination physics, these data favor the presence of a peak in the effective DE density at $z\sim 0.5$ at the $\sim 2.5-3\sigma$ CL, implying a crossing of the phantom divide from the phantom to the quintessence regime. This feature has been corroborated in several studies and from different perspectives \cite{Chakraborty:2024xas,Gomez-Valent:2024tdb,Benisty:2024lmj,Poulin:2024ken,Ye:2024ywg,Wolf:2024eph,Wolf:2024stt,Park:2024vrw,Gomez-Valent:2024ejh,Seto:2024cgo,Toda:2024ncp,Odintsov:2024woi,Giare:2025pzu,Keeley:2025stf,Khoury:2025txd,Pan:2025psn,Lu:2025gki,Wolf:2025jed,Chaussidon:2025npr,Chakraborty:2025syu,Yang:2025mws,Chen:2025wwn,Giani:2025hhs,Cai:2025mas,Braglia:2025gdo,Ozulker:2025ehg,Poulin:2025nfb,Camarena:2025upt,Gomez-Valent:2025mfl,Wang:2025znm,Yang:2025uyv,Yao:2025wlx,Nojiri:2025low,Artola:2025zzb,Mishra:2025goj,Wang:2025znm,Goh:2025upc,Tsujikawa:2025wca,Wolf:2025acj,SanchezLopez:2025uzw,Adi:2025hyj,Alestas:2025syk,Sharma:2025iux,Efstratiou:2025iqi,Cheng:2025yue,Ghedini:2025epp,deCruzPerez:2025dni,Toda:2025dzd,Toda:2025kcq,Li:2026xaz,Ibarra-Uriondo:2026zbp,Akarsu:2026anp,Park:2026iqa,Jhaveri:2026bla,Wang:2026wrk,Gomez-Valent:2026ept,Artola:2026tgs,Wang:2026vqw}, including model-independent reconstruction approaches \cite{DESI:2024aqx,Jiang:2024xnu,DESI:2025fii,Berti:2025phi,Li:2025ops,Gonzalez-Fuentes:2025lei,Gonzalez-Fuentes:2026rgu}.

Nevertheless, late-time dynamical DE in the form of a self-conserved component is clearly insufficient to alleviate some other longstanding tensions and anomalies within the $\Lambda$CDM model, such as the growth tension. In particular, data from redshift-space distortions (RSD) and weak gravitational lensing typically favor a lower level of structure formation than that predicted not only by the best-fit $\Lambda$CDM model from the Planck analysis, but also by the quintom scenario mentioned above \cite{Macaulay:2013swa,Gomez-Valent:2017idt,Nesseris:2017vor,Gomez-Valent:2018nib,Wright:2020ppw,Benisty:2020kdt,Nunes:2021ipq,DES:2021wwk,Miyatake:2023njf,Nguyen:2023fip,Toda:2024fgv,Semenaite:2025ohg} (see, however,  \cite{Wright:2025xka,Du:2026cly}). Although a larger neutrino mass ~\cite{Toda:2024uff} or interactions in the dark sector (see, e.g., \cite{Gomez-Valent:2018nib,Barros:2018efl,SolaPeracaula:2021gxi,Poulin:2022sgp,Sabogal:2024yha,deCruzPerez:2025dni}) can reduce the tension at some extent, it could also be an indication of modified gravity (MG) effects at cosmological scales.

In this {\it Letter}, we investigate such a possibility considering both dynamical DE through the Chevallier-Polarski-Linder (CPL) form of the equation-of-state (EoS) parameter \cite{Chevallier:2000qy,Linder:2002et}, $w(a)=w_0+w_a(1-a)$, -- which is flexible enough to mimic the behavior of a wide spectrum of models at the background level -- and by binning the effective gravitational coupling to parametrize the departures from General Relativity (GR) at the perturbations level -- which has also been shown to be a very robust way of compressing MG effects \cite{Denissenya:2017uuc,Toda:2024fgv}. We perform the binning not only in redshift, but also in scale, and show that this is crucial to accommodate the RSD data without jeopardizing the correct description of the CMB power spectra. We demonstrate that the MG effects should enter as a suppression of power below $z\sim 1$ at scales $\lambda_c\lesssim \mathcal{O}(150)$ Mpc, with the best-fit value at $\lambda_c\sim 40$ Mpc  -- well below the characteristic scale explored by the DESI collaboration in \cite{Ishak:2024jhs} -- and that these effects strengthen the signal of physics beyond $\Lambda$CDM, despite the obvious enlargement of the parameter space.


\section{Binning strategy}\label{sec:binning_method}
Modified gravity effects will in general impact the cosmological clustering of matter by means of an effective gravitational coupling \begin{equation}
G_{{\rm eff}}(k,a)=\mu(k,a)G\,, 
\end{equation}
which can deviate from the Newton constant $G$ in different cosmic epochs and length scales. It controls how density perturbations are linked to the amplitude of the scalar potential in the Newtonian gauge, $\psi$, through the modified Poisson equation, which in momentum space reads \cite{Pogosian:2010tj},
\begin{equation}
k^{2}\psi(k,a)  =-4\pi G_{{\rm eff}}(k,a)a^{2}\sum_{i}\rho_{i}(a)\Delta_{i}(k,a)\,,
\end{equation}
where $\rho_i$ and $\Delta_i$ are the background energy densities and the gauge-invariant density contrasts of the various species $i$, respectively.

Instead of adopting a concrete MG theory to compute $G_{\rm eff}(k,a)$, we employ a phenomenological parametrization in order to learn its optimal shape. We approximate $G_{\rm eff}$ at small wavelengths (i.e., at $k\gg k_c$)  as a step function in the scale factor (or redshift), using the binning strategy for $\mu$ already tested in \cite{Denissenya:2017uuc,Toda:2024fgv},
\begin{equation}\label{eq:2p}
\mu(a,k\gg k_c)=
    \begin{cases}
        \mu_1 & \text{if }\,a\leq0.5\quad (z\leq 1),\, 
        \\
        \mu_2 & \text{if } \,0.25\leq a<0.5 \quad(1<z\leq 3),\,
    \\
        1 & \text{if }\,a<0.25 \quad (z >3)\,.
    \end{cases}
\end{equation}
To ensure the continuity of the $\mu$ function, we smooth the step function using hyperbolic tangents.
In addition, to prevent anomalous behavior -- in the form of an excess of power -- in the low-$\ell$ region of the CMB power spectrum, we force $G_{\rm eff}\to G$ at large wavelengths (i.e. at $k\ll k_c$). Thus, we consider the function,

\begin{equation}\label{eq:mu}
\begin{aligned}\mu(z,k)= 1+\frac{1}{2}&\left[\mu_{2}+\frac{1}{2}(\mu_{1}-\mu_{2})\left(1+\tanh\!\left[\beta\left(1-z\right)\right]\right)-1\right]\\
 & \times\left(1+\tanh\!\left[\beta\left(3-z\right)\right]\right)\frac{k^{2}}{k^2+k_c^2},
\end{aligned}
\end{equation}
and take $\beta=5$ to ensure a fast enough transition in redshift (as in \cite{Toda:2024fgv}). We note that $k_c$ is precisely the wavenumber at which for a given redshift the deviation from GR reaches half of its maximum value; $\lambda_c = 2\pi/k_c$ is the corresponding comoving length, to be determined by the data.

In this work, we constrain MG at purely linear scales, thus remaining agnostic about the screening mechanisms required to recover GR locally, which would operate in the non-linear regime. It is also important to mention that, by construction, we retrieve GR at all scales at $z>3$, so the bounds on $G_{\rm eff}$ coming from big bang nucleosynthesis \cite{Uzan:2024ded} and CMB effects prior to photon decoupling are of course trivially fulfilled. These constraints could, otherwise, play a non-negligible role (see, e.g., \cite{Avilez:2013dxa,Sola:2015wwa,Alvey:2019ctk,SolaPeracaula:2020vpg}).

As already mentioned in the Introduction, we model the background MG effects with a self-conserved dynamical dark energy term in the CPL form.


\begin{table}
\begin{center}
\begin{tabular}{| c | c |c | c |}
\multicolumn{1}{c}{Survey} &  \multicolumn{1}{c}{$z$} &  \multicolumn{1}{c}{$f(z)\sigma_{12}(z)$} & \multicolumn{1}{c}{{\small References}}
\\\hline
ALFALFA & $0.013$ & $0.46\pm 0.06$ & \cite{Avila:2021dqv}
\\\hline
6dFGS+SDSS & $0.035$ & $0.338\pm 0.027$ & \cite{Said:2020epb}
\\\hline
GAMA & $0.18$ & $0.29\pm 0.10$ & \cite{Simpson:2015yfa}
\\ \cline{2-4}& $0.38$ & $0.44\pm0.06$ & \cite{Blake:2013nif}
\\\hline
 WiggleZ & $0.22$ & $0.42\pm 0.07$ & \cite{Blake:2011rj} \tabularnewline
\cline{2-3} & $0.41$ & $0.45\pm0.04$ & \tabularnewline
\cline{2-3} & $0.60$ & $0.43\pm0.04$ & \tabularnewline
\cline{2-3} & $0.78$ & $0.38\pm0.04$ &
\\\hline
DR12 BOSS & $0.32$ & $0.427\pm 0.056$  & \cite{Gil-Marin:2016wya}\\ \cline{2-3}
 & $0.57$ & $0.426\pm 0.029$ &
\\\hline
VIPERS & $0.60$ & $0.49\pm 0.12$ & \cite{Mohammad:2018mdy}
\\ \cline{2-3}& $0.86$ & $0.46\pm0.09$ &
\\\hline
VVDS & $0.77$ & $0.49\pm0.18$ & \cite{Guzzo:2008ac},\cite{Song:2008qt}
\\\hline
FastSound & $1.36$ & $0.482\pm0.116$ & \cite{Okumura:2015lvp}
\\\hline
eBOSS Quasar & $1.48$ & $0.462\pm 0.045$ & \cite{eBOSS:2020gbb}
\\\hline
 \end{tabular}
\end{center}
\caption{RSD data. See the quoted works for further details.}
\label{tab:fs8_table}
\end{table}

\section{Data and Methodology}\label{sec:data}
We implement Eq. \eqref{eq:2p} in the Einstein-Boltzmann code \texttt{MGCAMB}~\cite{Zhao:2008bn,Hojjati:2011ix,Zucca:2019xhg,Wang:2023tjj} -- which is the modified version of \texttt{CAMB}~\cite{Lewis:1999bs,Howlett:2012mh} for modified gravity -- and perform a Monte Carlo Markov chain (MCMC) analysis of the model using the public code \texttt{Cobaya}~\cite{Torrado:2020dgo},
requiring the Gelman-Rubin convergence criterion \cite{GelmanRubin} $R - 1 < 0.03$. 

In addition to the six standard cosmological parameters shared with the $\Lambda$CDM model, we sample the MG parameters $\mu_1, \mu_2 \in [0.0, 2.0]$. We first allow $\lambda_c^2 \in [4\times10^2\,{\rm Mpc}^2, 4\times10^6\,{\rm Mpc}^2]$
to vary freely in the MCMC analysis. This interval corresponds to
the original prior $k_c^{-2}\in[10,10^5]\,{\rm Mpc}^2$, after using
$\lambda_c=2\pi/k_c$.
We consider two different scenarios, assuming either a $\Lambda$CDM or a CPL background, with $w_0 \in [-1.3, -0.5]$ and $w_a \in [-2.0, 1.0]$ in the latter case. Regarding the neutrino sector, we assume the minimal-mass normal hierarchy, corresponding to $m_\nu = (0, 0, 0.06)\,\mathrm{eV}$.

\begin{figure}[t!]
    \centering
    \includegraphics[scale=0.4]{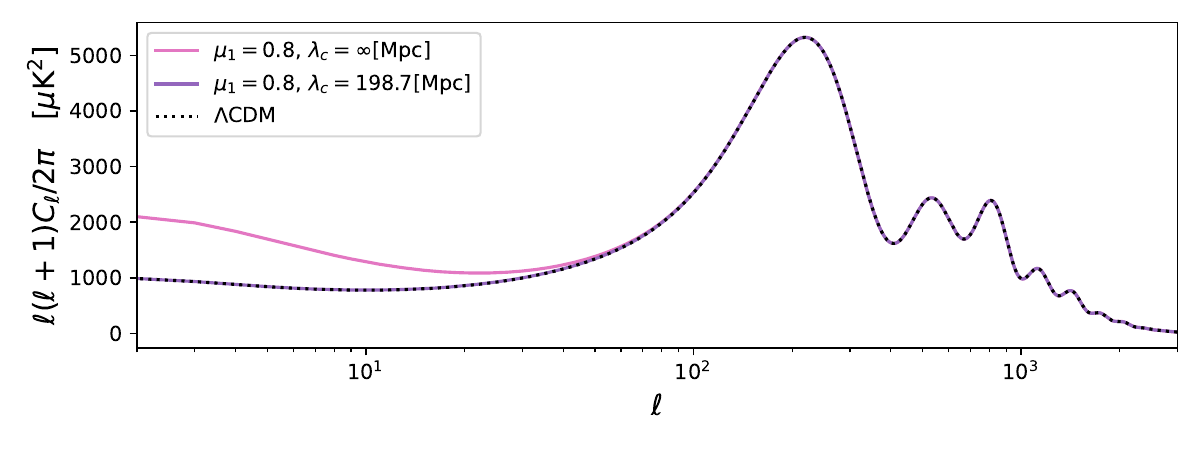}
    \includegraphics[scale=0.45]{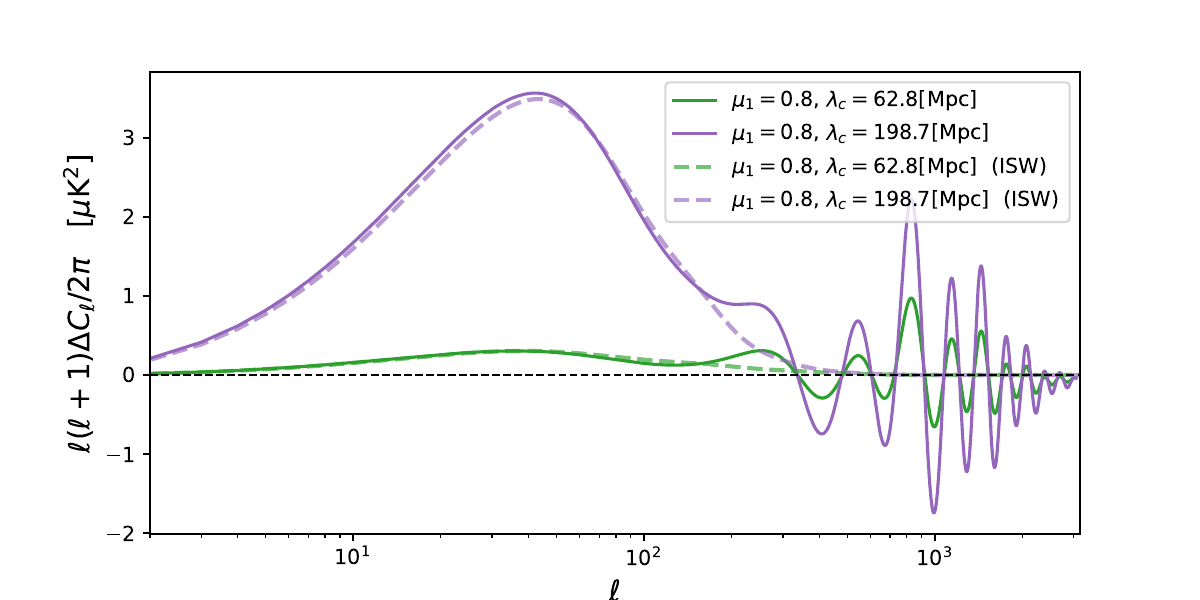}
    \caption{{\it Upper plot:} The CMB temperature power spectra for the standard $\Lambda$CDM model (with $\mu_1=\mu_2=1$) and two MG models with a $\Lambda$CDM background and $\mu_1=0.8$; {\it Lower plot}: The relative differences of the latter with the standard model. The dashed curves show the late-time ($z>29$) ISW effect on the CMB temperature power spectrum.}
    \label{fig:TT_mu1}
\end{figure}

We use the following datasets: \textbf{Planck CMB}, including the high-$\ell$ temperature and polarization likelihoods from \textsc{CamSpec} based on \texttt{NPIPE} (Planck PR4) data~\cite{Planck:2019nip,Rosenberg:2022sdy}, together with the low-$\ell$ likelihoods from \texttt{Commander} and \texttt{SimAll}; \textbf{Type Ia Supernovae} from the \texttt{Pantheon+} compilation~\cite{Brout:2022vxf}; \textbf{BAO} distance measurements from DESI DR2~\cite{DESI:2025zgx}; and \textbf{RSD} data. As argued in \cite{Sanchez:2020vvb,Forconi:2025cwp} (see also \cite{eBOSS:2021poy,Gomez-Valent:2021cbe,Semenaite:2022unt,Gomez-Valent:2022bku,Gomez-Valent:2023hov,Semenaite:2025ohg}), the use of $\sigma_8$ -- which corresponds to the rms mass fluctuations within spheres of radius $R_8 = 8h^{-1}\,\mathrm{Mpc}$ -- can introduce significant biases due to the explicit dependence of $R_8$ on $h$, especially in models predicting values of $H_0$ that differ substantially from those typically obtained in $\Lambda$CDM \cite{Forconi:2025cwp}. For this reason, we instead employ RSD data in terms of $f\sigma_{12}$ (cf. Table~\ref{tab:fs8_table}), where $\sigma_{12}$ denotes the rms mass fluctuations at the scale $R_{12}=12\,\mathrm{Mpc}$, which is independent of $h$. 
In CAMB, following the Planck paper~\cite{Planck:2015fie}, the value of $f\sigma_R$ is computed from the velocity--density correlation as
\begin{equation}\label{eq:fsigmaR}
f\sigma_R(z) = \left(\sigma^{(v\delta)}_R(z)\right)^2/\sigma^{(\delta\delta)}_R(z)\,.
\end{equation}
Here, 
\begin{equation}
\sigma^{(v\delta)}_R(z)=\left[\int \frac{dk}{2\pi^2} k^2 P_{v\delta}(k,z) W^2(kR)\right]^{1/2}
\end{equation}
denotes the velocity--density correlation, with $P_{v\delta}(k,z)=f(k,z)P_{\delta\delta}(k,z)$, $f(k,z)=d\ln\delta_m/d\ln a$ the growth rate and $\delta_m(k,z)$ the matter density contrast. $W(kR)$ is the Fourier transform of a spherical top-hat window function of radius $R$. The function  $\sigma^{(\delta\delta)}_R\equiv\sigma_R$ is, instead, the matter fluctuation amplitude~\cite{Lewis_CAMB_Documentation_2026}. For models with a $k$-dependent growth rate on the linear scales probed by galaxy surveys, such as the one studied here, it is essential to use a theoretical expression for the RSD observable that correctly relates the theory to the measured quantity, since galaxy surveys assume that $f$ is scale-independent. The growth rate $f$ entering Eq. \eqref{eq:fsigmaR} is actually an effective one, weighted over the linear scales allowed by the window function. Appendix A is devoted to demonstrating the validity of the definition employed in this work and comparing it with an alternative one, which yields fully consistent results.

\begin{figure}[t!]
    \centering
    \includegraphics[scale=0.42]{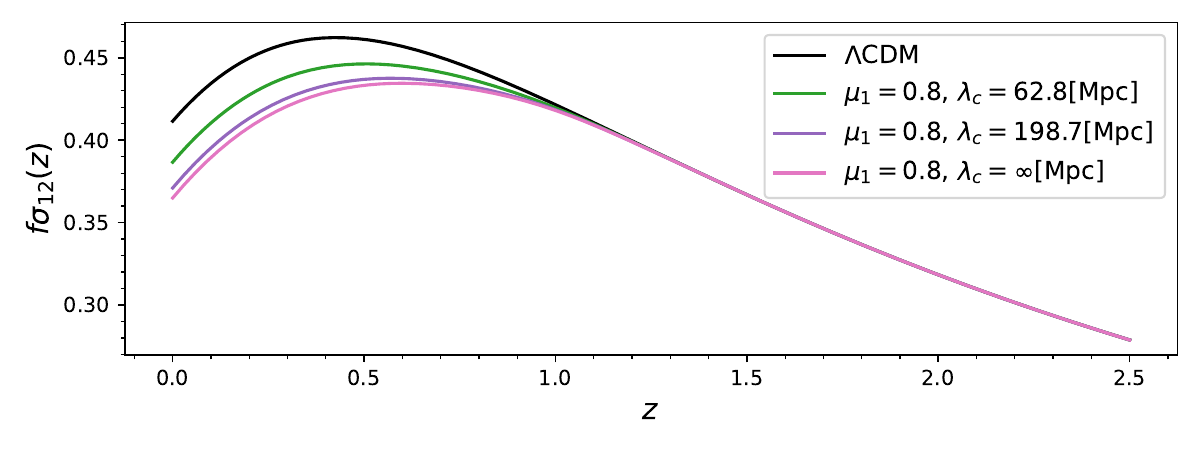}
    \includegraphics[scale=0.42]{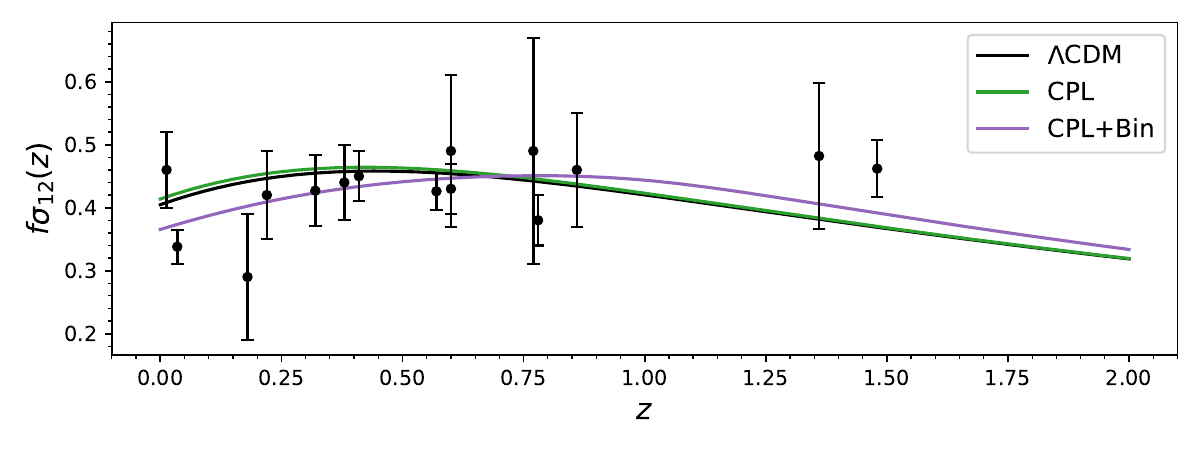}
    \caption{{\it Upper plot}: As in Fig. \ref{fig:TT_mu1}, but for $f\sigma_{12}(z)$. {\it Lower plot:} The best-fit curves for the models with  $\Lambda$CDM and CPL background and standard gravity (in black and green, respectively), and the CPL with MG and  varying $\lambda_c$ (in purple).}
    \label{fig:fsigma}
\end{figure}


\section{Results \& Discussion}\label{sec:results}

\begin{table*}[t]
\centering {\footnotesize{}%
\begin{tabular}{|lccccc|}
\hline
{\footnotesize Parameter } & {\footnotesize$\Lambda$CDM} & {\footnotesize Bin (varying $\ensuremath{\lambda_{c}}$)} & {\footnotesize Bin ($\ensuremath{\ensuremath{\lambda_{c}}=62.8}$ Mpc)}  & {\footnotesize Bin ($\ensuremath{\ensuremath{\lambda_{c}}=198.5}$ Mpc)}  & {\footnotesize Bin ($\ensuremath{\lambda_{c}=\infty}$) }\tabularnewline
\hline 
{\footnotesize{\footnotesize{}{\footnotesize\boldmath}$\mu_{1}${\footnotesize}} } & {\footnotesize$1$} & {\footnotesize$<0.72$} & {\footnotesize$\ensuremath{0.56^{+0.26}_{-0.34}}$} & {\footnotesize$\ensuremath{0.69\pm0.17}$} & {\footnotesize$\ensuremath{1.025^{+0.035}_{-0.045}}$}\tabularnewline
{\footnotesize{\footnotesize{}{\footnotesize\boldmath}$\mu_{2}${\footnotesize}} } & {\footnotesize$1$} & {\footnotesize$0.96^{+0.39}_{-0.22}$} & {\footnotesize$\ensuremath{1.12^{+0.22}_{-0.19}}$} & {\footnotesize$\ensuremath{1.16\pm0.11}$} & {\footnotesize$\ensuremath{1.010\pm0.023}$}\tabularnewline
{\footnotesize{\footnotesize{}{\footnotesize\boldmath}$\log_{10}\ensuremath{\lambda_{c}}${\footnotesize}} } & {\footnotesize$-$} & {\footnotesize$<1.64$ ($<2.26\,\,95\%$)} & {\footnotesize 1.80} & {\footnotesize 2.30} & {\footnotesize$\infty$}\tabularnewline
{\footnotesize$H_{0}$ } & {\footnotesize$\ensuremath{68.32\pm0.28}$} & {\footnotesize$68.18\pm0.30$} & {\footnotesize$\ensuremath{68.15\pm0.29}$} & {\footnotesize$\ensuremath{68.24\pm0.30}$} & {\footnotesize$\ensuremath{68.41\pm0.30}$}\tabularnewline
{\footnotesize$\Omega_{\mathrm{m}}$ } & {\footnotesize$\ensuremath{0.3005\pm0.0036}$} & {\footnotesize$0.3024\pm0.0038$} & {\footnotesize$\ensuremath{0.3028\pm0.0038}$} & {\footnotesize$\ensuremath{0.3016\pm0.0039}$} & {\footnotesize$\ensuremath{0.2995\pm0.0037}$}\tabularnewline
{\footnotesize$f\sigma_{8}$ } & {\footnotesize$\ensuremath{0.4100\pm0.0046}$} & {\footnotesize$0.380\pm0.019$} & {\footnotesize$\ensuremath{0.370^{+0.019}_{-0.023}}$} & {\footnotesize$\ensuremath{0.373\pm0.022}$} & {\footnotesize$\ensuremath{0.4157^{+0.0089}_{-0.010}}$}\tabularnewline
{\footnotesize$f\sigma_{12}$ } & {\footnotesize$\ensuremath{0.4032\pm0.0053}$} & {\footnotesize$0.375\pm0.018$} & {\footnotesize$\ensuremath{0.365^{+0.019}_{-0.022}}$} & {\footnotesize$0.367\pm0.021$} & {\footnotesize$\ensuremath{0.4084^{+0.0089}_{-0.0099}}$}\tabularnewline
\hline 

{\footnotesize Parameter} & {\footnotesize CPL} & {\footnotesize CPL+Bin (varying $\ensuremath{\lambda_{c}}$)} & {\footnotesize CPL+Bin ($\ensuremath{\ensuremath{\lambda_{c}}=62.8}$ Mpc)}  & {\footnotesize CPL+Bin ($\ensuremath{\ensuremath{\lambda_{c}}=198.5}$ Mpc)}  & {\footnotesize CPL+Bin ($\ensuremath{\lambda_{c}=\infty}$)}\tabularnewline
\hline 
{\footnotesize{\footnotesize{}{\footnotesize\boldmath}$\mu_{1}${\footnotesize}}} & {\footnotesize$ 1 $} & {\footnotesize$ < 0.64 $} & {\footnotesize$\ensuremath{0.51^{+0.23}_{-0.33}}$} & {\footnotesize$\ensuremath{0.67^{+0.16}_{-0.18}}$} & {\footnotesize$\ensuremath{1.013^{+0.034}_{-0.044}}$}\tabularnewline
{\footnotesize{\footnotesize{}{\footnotesize\boldmath}$\mu_{2}${\footnotesize}}} & {\footnotesize$ 1 $} & {\footnotesize$ 0.90^{+0.39}_{-0.22} $} &{\footnotesize$\ensuremath{1.08^{+0.21}_{-0.17}}$} & {\footnotesize$\ensuremath{1.12^{+0.12}_{-0.11}}$} & {\footnotesize$\ensuremath{1.001^{+0.022}_{-0.025}}$}\tabularnewline
{\footnotesize{\footnotesize{}{\footnotesize\boldmath}$\log_{10}\ensuremath{\lambda_{c}}${\footnotesize}} } & {\footnotesize$-$} & {\footnotesize$<1.66$ ($<2.14\,\,95\%$)} & {\footnotesize 1.80} & {\footnotesize 2.30} & {\footnotesize$\infty$}\tabularnewline
{\footnotesize{\footnotesize{}{\footnotesize\boldmath}$w_{0}${\footnotesize}}} & {\footnotesize$\ensuremath{-0.857\pm 0.054} $} &{\footnotesize$ -0.841\pm 0.056 $} & {\footnotesize$\ensuremath{-0.839\pm0.055}$} & {\footnotesize$\ensuremath{-0.841\pm0.055}$} & {\footnotesize$\ensuremath{-0.861\pm0.054}$}\tabularnewline
{\footnotesize{\footnotesize{}{\footnotesize\boldmath}$w_{a}${\footnotesize}}} & {\footnotesize$\ensuremath{-0.50\pm 0.20} $} &{\footnotesize$ -0.61\pm 0.22 $} & {\footnotesize$\ensuremath{-0.61^{+0.23}_{-0.20}}$} & {\footnotesize$-0.58^{+0.22}_{-0.20}$} & {\footnotesize$\ensuremath{-0.48^{+0.22}_{-0.20}}$}\tabularnewline
{\footnotesize$H_{0}$} & {\footnotesize$\ensuremath{67.63\pm 0.60} $} & {\footnotesize$ 67.55\pm 0.60 $} &{\footnotesize$\ensuremath{67.50\pm0.60}$} & {\footnotesize$\ensuremath{67.48\pm0.60}$} & {\footnotesize$\ensuremath{67.62\pm0.59}$}\tabularnewline
{\footnotesize$\Omega_{\mathrm{m}}$} & {\footnotesize$\ensuremath{0.3084\pm 0.0056}  $} & {\footnotesize$ 0.3106\pm 0.0058 $} &{\footnotesize$\ensuremath{0.3112\pm0.0057}$} & {\footnotesize$\ensuremath{0.3109\pm0.0058}$} & {\footnotesize$\ensuremath{0.3082\pm0.0057}$}\tabularnewline
{\footnotesize$f\sigma_{8}$} & {\footnotesize$\ensuremath{0.4159\pm 0.0053} $} & {\footnotesize$ 0.380^{+0.019}_{-0.017} $} & {\footnotesize$\ensuremath{0.370^{+0.019}_{-0.022}}$} & {\footnotesize$\ensuremath{0.373\pm0.021}$} & {\footnotesize$\ensuremath{0.4179\pm0.0092}$}\tabularnewline
{\footnotesize$f\sigma_{12}$} & {\footnotesize$\ensuremath{0.4118\pm 0.0061} $} & {\footnotesize$ 0.377\pm 0.018 $} &{\footnotesize$\ensuremath{0.367^{+0.019}_{-0.022}}$} & {\footnotesize$0.370\pm0.021$} & {\footnotesize$\ensuremath{0.4138\pm0.0097}$}\tabularnewline
\hline 
\end{tabular}}\caption{Marginalized constraints at 68\% CL for the binning method either allowing $\lambda_c$ to vary in the MCMC analysis or fixing it to concrete values, assuming a $\Lambda$CDM
or a CPL background. Our dataset includes data on CMB
(Planck PR4), BAO (DESI DR2), Type Ia supernovae
(Pantheon+) and RSD. $H_0$ is given in km/s/Mpc, $\lambda_c$ in Mpc, and the derived values of $f\sigma_8$ and $f\sigma_{12}$ (see Eq. \eqref{eq:fsigmaR}) are computed at $z=0$.}
\label{tab:Avaried} 
\end{table*}

Figure ~\ref{fig:TT_mu1} shows the CMB temperature power spectra for several values of $\mu_{1}$ and $\lambda_c$ (setting $\mu_2=1$).
At large scales ($\ell \lesssim 200$), when we assume a weaker effective gravitational constant, e.g. $\mu_{1}=0.8$, the late-time Integrated Sachs–Wolfe (ISW) effect is enhanced. This is because the ISW contribution to the temperature anisotropy is given by
$\int_{t_{L}}^{t_{0}} dt \left( \dot{\phi} + \dot{\psi} \right),$
where $t_L$ and $t_0$ are the cosmic times at the last scattering and today, respectively. A smaller effective gravitational constant modifies the time evolution of the gravitational potentials, leading to a larger integrated contribution.
In the limit $\lambda_c\to\infty\,\mathrm{Mpc}$, in which the effective gravitational constant is suppressed at all scales with respect to GR, the low-$\ell$ power spectrum becomes significantly enhanced. Instead, when a scale-dependent suppression is introduced, setting e.g. $\lambda_c = 62.8\,\mathrm{Mpc}$ or $\lambda_c = 198.5\,\mathrm{Mpc}$, the enhancement in the power spectrum is reduced to only a few $\mu\mathrm{K}^2$ at most in the entire multipole range. The remaining differences can be attributed to two effects. 
The first is still the ISW effect, whose contribution is visible at multipoles $\ell < 200$, as indicated by the dashed lines in the lower plot of Fig. \ref{fig:TT_mu1}. The second arises from the suppression of gravitational lensing. Lensing smooths the acoustic peaks and when it is reduced due to a weaker effective gravitational coupling, the smoothing   becomes less efficient, resulting in sharper acoustic features in the high-$\ell$ region.

The effect of the MG parameters $\mu$ on the RSD observable $f\sigma_{12}(z)$ is shown in Fig.~\ref{fig:fsigma}. Values of $\mu<1$ ($\mu>1$) suppress (enhance) the growth of structure relative to GR, and the impact becomes stronger for larger values of $\lambda_c$.

\begin{figure}
    \centering
    \includegraphics[scale=0.42]{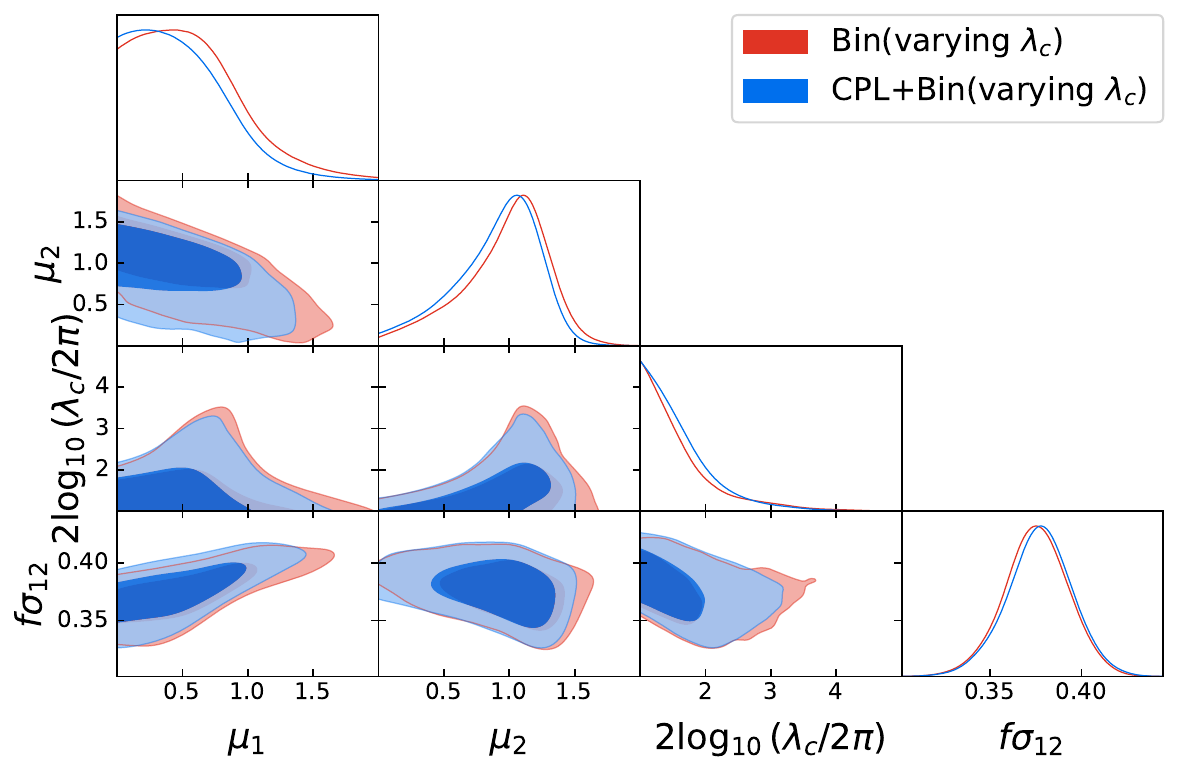}
    \caption{Triangle plots (at $68\%$ and $95\%$ CL) for the models with $\Lambda$CDM (in red) and CPL (in blue) backgrounds including MG effects, with $\lambda_c$ varied in the MCMC analysis. The corresponding parameter constraints are listed in Table~\ref{tab:Avaried}.}
    \label{fig:Avaried}
\end{figure}

We present the contour plots and one-dimensional posterior distributions for the MG parameters of the binning method with varying $\lambda_c$ in Fig.~\ref{fig:Avaried}, and list the corresponding set of parameter constraints in Table~\ref{tab:Avaried}. 
Assuming a $\Lambda$CDM background, we obtain an upper bound on the critical wavelength of $\lambda_c < 182\,\mathrm{Mpc}$ (95\% CL), which becomes tighter when a dynamical dark energy background is considered. 
Furthermore, as $\lambda_c$ decreases, the constraints on $\mu$ become weaker, as it is clear from the analyses with fixed $\lambda_c$ reported in Table \ref{tab:Avaried}. This is because the impact of $\mu$ on the growth rate is increasingly suppressed at smaller $\lambda_c$, as illustrated in Fig.~\ref{fig:fsigma}.

\begin{figure*}
    \centering
    \includegraphics[scale=0.42]{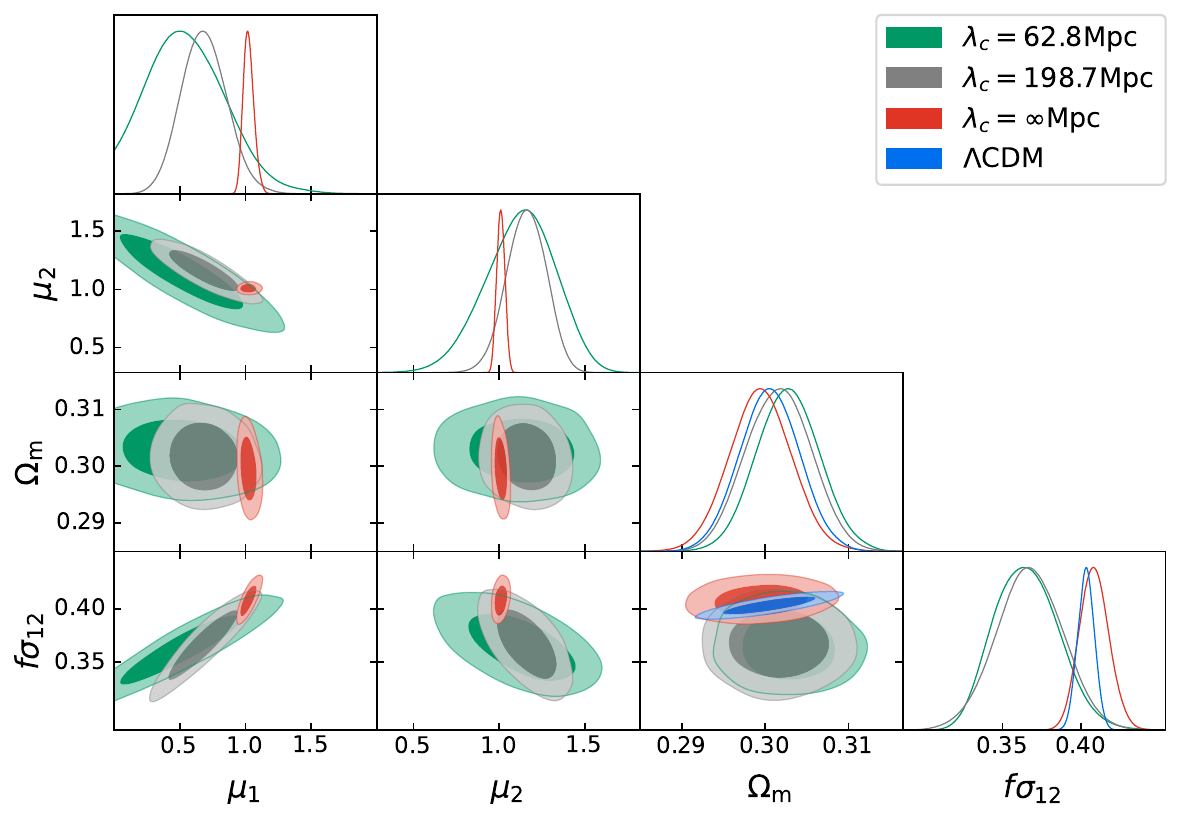}
    \includegraphics[scale=0.42]{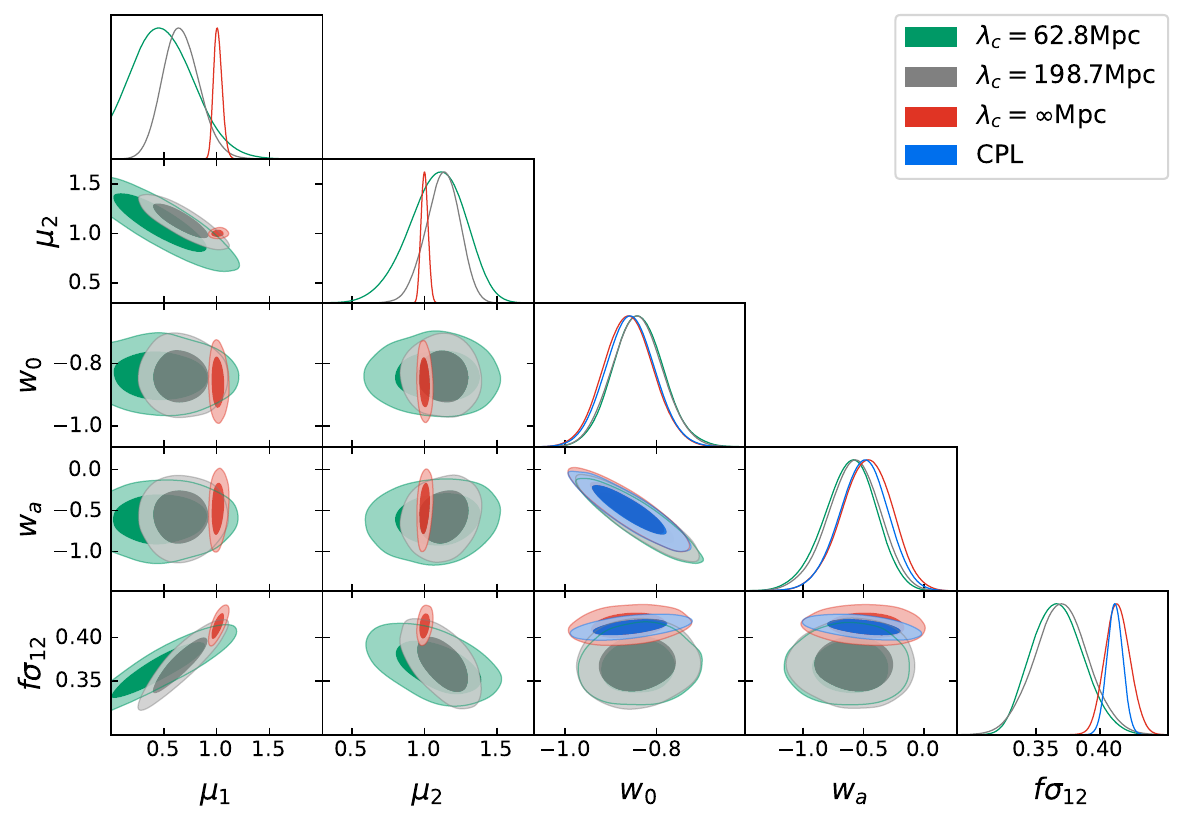}
    \caption{The same as in Fig. \ref{fig:Avaried}, but setting $\lambda_c$ to concrete values. We provide results with a $\Lambda$CDM and a CPL background (on the left and right, respectively).}
    \label{fig:Bin}
\end{figure*}

We also examine whether changes in the background expansion affect the constraints on the bin parameters. The region of parameter space preferred by the data in the CPL model with standard gravity leads to a larger growth rate compared to $\Lambda$CDM (cf. again Table \ref{tab:Avaried} and the bottom plot of Fig. \ref{fig:fsigma}). This is essentially caused by the reduced amount of dark energy at $z\gtrsim 1$ relative to the standard model (see, e.g., Fig 4 in \cite{Gonzalez-Fuentes:2026rgu}), which is required to counterbalance the enhancement at low redshifts and reproduce the correct distance to the last-scattering surface. To compensate for this effect and explain the low measurements of $f\sigma_{12}(z)$ at $z<1$, slightly smaller values of $\mu_1$ (lower than 1) are favored when the CPL parameters are allowed to vary. However, the shift induced by the background evolution is of the order  $\lesssim 10^{-1}$ and remains much smaller than the variation caused by the choice of $\lambda_c$. On the other hand, the two highest-redshift RSD data points (at $z>1$) lie above the best-fit $\Lambda$CDM curve, which implies that values of $\mu_2\gtrsim1$ are not disfavored. The constraints on this parameter are, however, still fully compatible with unity at the $1\sigma$ CL. Since the background evolution in the best-fit CPL model already leads to enhanced matter growth --driven by the larger values of $\Omega_{\rm m}(z)$ -- the MG effects become less necessary. This explains why $\mu_2$ is mildly smaller in the CPL case than in $\Lambda$CDM, which is more evident in the analyses with fixed $\lambda_c$.

Motivated by the 68\% and 95\% upper bounds on $\lambda_c$ reported in the second column of Table~\ref{tab:Avaried}, we perform the analysis by fixing $\lambda_c$ to three representative values: $\lambda_c = 62.8\,\mathrm{Mpc}$, $\lambda_c = 198.5\,\mathrm{Mpc}$, and $\lambda_c = \infty\,\mathrm{Mpc}$. The corresponding results are displayed in Table \ref{tab:Avaried} and Fig.~\ref{fig:Bin}. As mentioned above, the value of $\lambda_c$ plays a crucial role in determining the constraints on $\mu$. Smaller values of $\lambda_c$ lead to weaker constraints on $\mu$. The constraints on $\mu_2$ show a similar trend, with $\mu_1$ and $\mu_2$ exhibiting a negative correlation, as expected. 
For example, assuming a $\Lambda$CDM background, we obtain $\mu_1 = 0.56^{+0.26}_{-0.34}$ for $\lambda_c = 62.8\,\mathrm{Mpc}$, $\mu_1 = 0.69 \pm 0.17$ for $\lambda_c = 198.5\,\mathrm{Mpc}$, and $\mu_1 = 1.025^{+0.035}_{-0.045}$ for $\lambda_c = \infty\,\mathrm{Mpc}$. The constraints on $\mu_2$, instead, tend to favor the region above $\mu_2=1$, but are in any case compatible with the standard GR value at $1\sigma$ CL. 

It is important to emphasize that the use of RSD data based on $f\sigma_{12}$ instead of $f\sigma_8$ does not play a major role in our analysis, since the values of $H_0$ in both the $\Lambda$CDM and CPL models, with and without MG effects, remain close to the Planck/$\Lambda$CDM value \cite{Planck:2018vyg} (cf. Table~\ref{tab:Avaried}), which is the value typically adopted as the fiducial input in observational analyses. Consequently, the relative shifts in $f\sigma_R$ induced by choosing a smoothing scale of either $8h^{-1}\,\mathrm{Mpc}$ or $12\,\mathrm{Mpc}$ are at most $0.5\%$, well below the current sensitivity of RSD measurements \cite{Forconi:2025cwp}. This is also reflected in the similarity between the posterior values of $f\sigma_8(z=0)$ and $f\sigma_{12}(z=0)$ reported in Table~\ref{tab:Avaried}.

In Appendix B, we provide details on the statistical significance of the results displayed in Table~\ref{tab:Avaried}. We find that modifications of gravity on top of a $\Lambda$CDM background do not lead to a substantial decrease in the minimum $\chi^2$, $\chi^2_{\rm total}$, which is reduced by at most $\sim 5$ units when $\lambda_c$ is allowed to vary. Changing the background to CPL while keeping standard GR at the perturbation level instead decreases $\chi^2_{\rm total}$ by approximately 7 units, whereas the inclusion of MG effects further improves the fit, leading to a total reduction of about 15 units. When analyzed using the Planck+DESI DR2+PantheonPlus+RSD dataset combination, this model is preferred over the standard $\Lambda$CDM scenario at the $2.63\sigma$ confidence level and yields $\Delta$AIC$=5.45$ in favor of the CPL+MG scenario, after duly penalizing the use of additional parameters. Therefore, RSD data not only preserve the signal for new physics already present at the background level, but also strengthen it through the inclusion of MG effects. Replacing the \texttt{Pantheon+} SNIa compilation with DES-Dovekie further enhances the signal to $2.80\sigma$, cf. Appendix~C.


\section{Conclusions}\label{sec:conclusions}

We find that, in the binning method, which is an effective parametrization that compresses a wide class of modified gravity theories, deviations from standard gravity must be suppressed at late times on linear cosmological scales of  $\lesssim 150\,\mathrm{Mpc}$ (95\% CL). In this work, we also demonstrate that this scale is determined not only by the late-time ISW effect but also by the gravitational lensing effect on the CMB.

We examine the binning method assuming a $\Lambda$CDM background using the dataset CMB(Planck PR4)+BAO(DESI)+SNIa(PantheonPlus)+RSD and find
\begin{align}
&
\begin{cases}
\mu_1=0.56^{+0.26}_{-0.34}  \\
\mu_2 =1.12^{+0.22}_{-0.19}  \\
\end{cases}\,\,\,\,\,\,\,\,\,\,\,\mathrm{Bin} \,(\lambda_c=62.8 \,\mathrm{Mpc}), \\
&
\begin{cases}
\mu_1 = 0.69\pm 0.17  \\
\mu_2 = 1.16\pm 0.11  \\
\end{cases}\quad \mathrm{Bin} \,(\lambda_c=198.5 \,\mathrm{Mpc})\,.
\end{align}
Non-standard gravity theories ($\mu\neq1$) are preferred by appropriately suppressing large-scale modifications in the late universe, at $z\lesssim 1$. Strong departures from GR at low redshifts, such as those implied by our constraints, can arise, e.g., in scalar--tensor theories featuring a direct connection between $\mu$ and the dark energy dynamics (see, e.g., \cite{Wang:2023tjj,Wolf:2025jed} for models with a big low-$z$ enhancement of structure growth rather than a suppression, and \cite{Wittner:2020yfc} for a scenario with transient weak gravity).

We also examine whether changes in the background expansion affect the constraints on the bin parameters and find that they remain largely unchanged, but with mildly smaller uncertainties. In our parametrized framework, the required modifications in the linear perturbation sector are therefore robust under changes in the background, indicating that the effective dark energy dynamics and the necessary departures from standard gravity are effectively decoupled. The fact that the MG sector is largely independent of the expansion history is not problematic from a theoretical standpoint, since both sectors can be described by different functions within the effective field theory of dark energy \cite{Gubitosi:2012hu,Frusciante:2019xia}, which can in turn be mapped onto concrete Lagrangians, e.g. of scalar-tensor theories.

It is also useful to compare our findings with the recent DESI full-shape modified-gravity analysis~\cite{Ishak:2024jhs}. In that work, the scale dependence of the binned MG functions is introduced through a smooth hyperbolic-tangent transition at a fixed scale $k_c=0.01\,{\rm Mpc}^{-1}$. Following their notation, the redshift-and-scale binned $\mu$ function for $z<1$ and $1<z<2$ is written as
\[
\mu_{z_1}(k)=\frac{\mu_2+\mu_1}{2}
+\frac{\mu_2-\mu_1}{2}
\tanh\left(\frac{k-k_c}{k_{\rm tw}}\right),
\]
and
\[
\mu_{z_2}(k)=\frac{\mu_4+\mu_3}{2}
+\frac{\mu_4-\mu_3}{2}
\tanh\left(\frac{k-k_c}{k_{\rm tw}}\right)\,,
\]
respectively, with $k_{\rm tw}=k_c/10$. For a \(\Lambda\)CDM background, using DESI+Planck(PR3) +DESY3+DESY5SN, they obtained
$$
\mu_1=0.97\pm0.18, \,
\mu_2=0.95\pm0.11,\,
$$
$$
\mu_3=0.83\pm0.24,\,
\mu_4=1.14\pm0.15, \,
$$
which are all consistent with the GR prediction $\mu_i=1$ although their mean values for $k>k_c$ prefer a suppression of gravity at $z<1$ and an enhancement at $1<z<2$ (cf. their values of $\mu_2$ and $\mu_4$, respectively), in accordance with our findings. In addition, they obtained $\mu_2=0.86\pm 0.14$ and $\mu_4=1.02\pm0.15$ by considering dynamical dark energy at the background level. This is again in good agreement with our results. We note, however, that their transition wave mode is one order of magnitude smaller than the one leading to the lowest $\chi^2$ in our study, $k_c=2\pi/\lambda_c\simeq0.10\,{\rm Mpc}^{-1}$. Therefore, the two analyses should not be regarded as being in  tension; rather, they probe different scale ranges. Our results indicate that moving their transition scale to larger $k_c$ can lead to a mildly stronger preference for a departure from GR.

By allowing for MG effects, the signal for new physics is enhanced relative to the CPL scenario with standard gravity, even after properly penalizing the use of additional degrees of freedom. At the same time, the CPL background parameters remain fully consistent with those obtained in the GR analysis, although they are mildly shifted toward the quintom region. Replacing \texttt{Pantheon+} with DES-Dovekie leads to results that remain fully consistent for both the CPL and MG parameters. However, the statistical evidence for new physics is further enhanced with DES-Dovekie.

Although the statistical significance of the reported hints are still moderate, our results may indicate that scale-dependent modified gravity in a dynamical dark energy background is necessary to explain the suppressed growth of structure inferred from RSD observations at $z<1$, while keeping the good description of the low-redshift and CMB data.

 \section*{Acknowledgements}
 YT is supported by JSPS KAKENHI Grant No. 26K17154. AGV is funded by “la Caixa” Foundation (ID 100010434) and the European Union's Horizon 2020 research and innovation programme under the Marie Sklodowska-Curie grant agreement No 847648, with fellowship code LCF/BQ/PI23/11970027. He is also supported by projects PID2022-136224NB-C21 (MICIU), 2021-SGR-00249 (Generalitat de Catalunya) and CEX2024-001451-M (ICCUB). The authors acknowledge the participation in the COST Action CA21136 “Addressing observational tensions in cosmology with systematics and fundamental physics” (CosmoVerse).

 \appendix
\begin{table*}[t!]
\[
\begin{tabular}{lccccc}
\ensuremath{\Lambda}CDM background\\
\hline\hline Parameter  &  standard  &  Bin (varying \ensuremath{\lambda_{c}})  &  Bin (\ensuremath{\lambda_{c}}=62.8 Mpc)  &  Bin (\ensuremath{\lambda_{c}}=198.5 Mpc)  &  Bin (\ensuremath{\lambda_{c}}=\ensuremath{\infty})\\
\hline {\boldmath\ensuremath{\mu_{1}}}  &  1$^*$  &  0.10  &  0.45  &  0.69  &  1.02\\
 {\boldmath\ensuremath{\mu_{2}}}  &   1$^*$  &  1.19  &  1.17  &  1.12  &  1.01\\
 {\boldmath\ensuremath{\log_{10}\lambda_{c}}}  &  -  &  1.52  &  1.80$^*$  &  2.30$^*$  &  \ensuremath{\infty}$^*$\\
 \ensuremath{H_{0}}  &  68.37  &  68.12  &  68.11  &  68.20  &  68.46\\
 \ensuremath{f\sigma_{8}}  &  0.412  &  0.365  &  0.363  &  0.370  &  0.419\\
 \ensuremath{f\sigma_{12}}  &  0.405  &  0.361  &  0.358  &  0.365  &  0.411\\
\hline \ensuremath{\chi_{\mathrm{BAO}}^{2}}  &  11.08  &  12.58  &  12.56  &  12.06  &  10.75\\
 \ensuremath{\chi_{\mathrm{RSD}}^{2}}  &  20.39  &  12.48  &  12.53  &  13.55  &  22.84 \\
 \ensuremath{\chi_{\mathrm{Planck}}^{2}}  &  10967.87  &  10969.55  &  10970.52  &  10969.98  &  10965.17\\
 \ensuremath{\chi_{\mathrm{SN}}^{2}}  &  1406.07  &  1405.50  &  1405.48  &  1405.66  &  1406.29\\ \hline
 \ensuremath{\chi_{\mathrm{total}}^{2}}  &  12405.41 &  12400.12 &  12401.10 & 12401.25  & 12405.05
\\
$E_{\Lambda{\rm CDM}}$ & -  &  $1.44\sigma$ & $1.57\sigma$  & $1.54\sigma$   & $0.21\sigma$
\\
$\Delta{\rm AIC}$ & - & $-0.71$ & +0.31  & +0.16  & -3.64
\\\hline
\hline 
\\ 
CPL background \\
\hline\hline Parameter  &  standard  &  Bin (varying \ensuremath{\lambda_{c}})  &  Bin (\ensuremath{\lambda_{c}}=62.8 Mpc)  &  Bin (\ensuremath{\lambda_{c}}=198.5 Mpc)  &  Bin (\ensuremath{\lambda_{c}}=\ensuremath{\infty})\\
\hline {\boldmath\ensuremath{\mu_{1}}}  &  1$^*$  &  0.28 &  0.23 &  0.73  &  0.98\\
 {\boldmath\ensuremath{\mu_{2}}}  &  1$^*$  &  1.16 & 1.16  &  1.08  &  0.99\\
 {\boldmath\ensuremath{\log_{10}\lambda_{c}}}  &  -  & 1.64  & 1.80$^*$  &  2.30$^*$  &  \ensuremath{\infty}$^*$\\
 {\boldmath\ensuremath{w_{0}}}  &  -0.853  &  -0.843 & -0.863  &  -0.881  &  -0.864\\
 {\boldmath\ensuremath{w_{a}}}  &  -0.50  & -0.56  &  -0.52 &  -0.43  &  -0.49\\
 \ensuremath{H_{0}}  &  67.60  & 67.40  & 67.49 &  67.67  &  67.71\\
 \ensuremath{f\sigma_{8}}  &  0.417  & 0.369  & 0.343  &  0.380 &  0.412\\
 \ensuremath{f\sigma_{12}}  &  0.413  & 0.366 & 0.341  &  0.376  &  0.407\\
\hline \ensuremath{\chi_{\mathrm{BAO}}^{2}}  &  9.13  & 9.49  & 10.28 &  9.74  &  9.40\\
 \ensuremath{\chi_{\mathrm{RSD}}^{2}}  &  21.38  & 12.35  & 11.83  &  14.01  &  19.74 \\
 \ensuremath{\chi_{\mathrm{Planck}}^{2}}  &  10964.26  & 10964.97  & 10967.74  &  10967.11  &  10965.90\\
 \ensuremath{\chi_{\mathrm{SN}}^{2}}  &  1402.93  & 1403.15  & 1402.85  &  1402.74  &  1402.82\\ \hline
 \ensuremath{\chi_{\mathrm{total}}^{2}}  & 12397.70  & 12389.96  & 12392.71  & 12393.59  & 12397.87
\\
$E_{\Lambda{\rm CDM}}$ & $2.31\sigma$ & $2.63\sigma$  & $2.48\sigma$  & $2.35\sigma$   & $1.60\sigma$
\\
$\Delta{\rm AIC}$ & +3.71 & +5.45 & +4.70 &  +3.82 & -0.46
\\\hline
\hline \end{tabular}
\]
\caption{The best-fit values of the cosmological parameters together with their corresponding
minimized chi-squared values, $\chi^2_{\rm total}$, exclusion level with respect to the $\Lambda$CDM, $E_{\Lambda{\rm CDM}}$, and difference of Akaike information criteria, $\Delta{\rm AIC}\equiv AIC_{\Lambda{\rm CDM}}-AIC_i$. The starred quantities have been fixed in the corresponding MCMC analyses. We note that $\chi^2_{\rm total}$ for the CPL with $\lambda_c=\infty$ is slightly larger (by 0.17 units) than in the standard CPL model. This is just due to small numerical inaccuracies in the determination of the
best-fit point. }\label{table-best}
\end{table*}

\section{$f\sigma_R$: from theory to the observable}

The theoretical expression employed in our analysis for the computation of the RSD observable $f\sigma_R$, given in Eq. \eqref{eq:fsigmaR}, can be decomposed as follows,

\begin{equation}
   f\sigma_{12}(z)\equiv f_{\rm eff}(z)\times\sigma^{(\delta\delta)}_{12}(z)\,, 
\end{equation}
with $\sigma^{(\delta\delta)}_{12}(z)$ defined as in the main text and $f_{\rm eff}(z)$ the effective growth rate,

\begin{figure}[t!]
    \centering
    \includegraphics[scale=0.5]{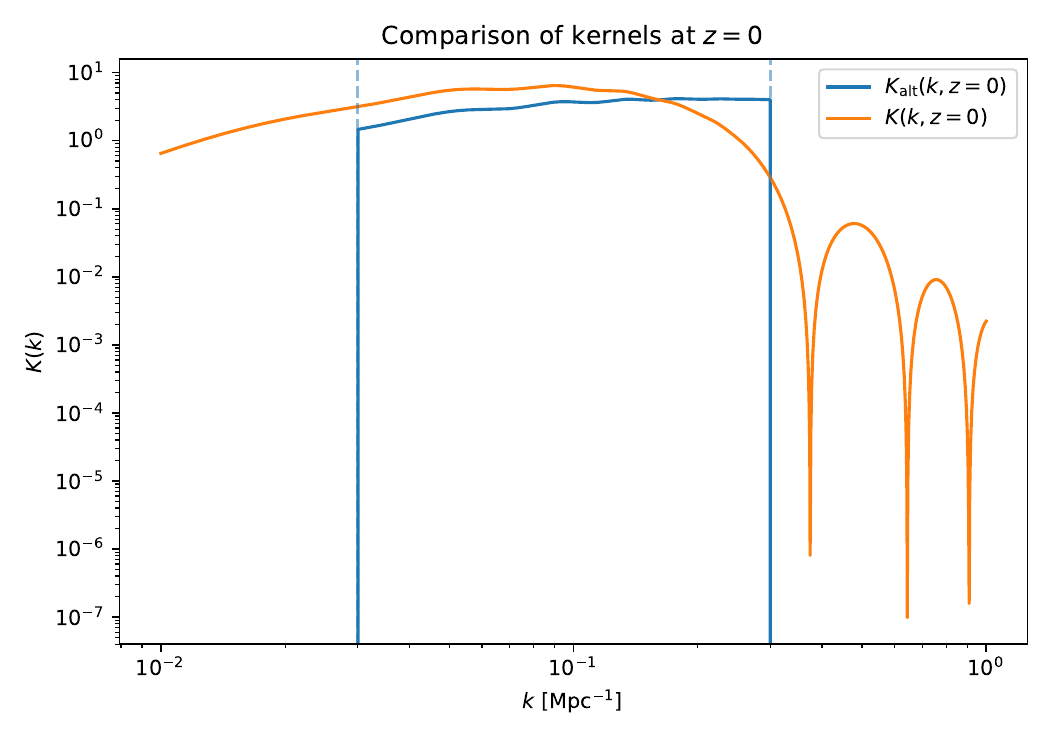}
    \caption{Comparison of the kernels \eqref{eq:kernel} and \eqref{eq:kernel2} entering the computation of the effective growth rates \eqref{eq:feff} and \eqref{eq:feff2}, respectively. The blue vertical dashed lines set the borders of the typical scale range probed by galaxy surveys with RSD.}
    \label{fig:kernel}
\end{figure}

\begin{figure}[t!]
    \centering
    \includegraphics[scale=0.42]{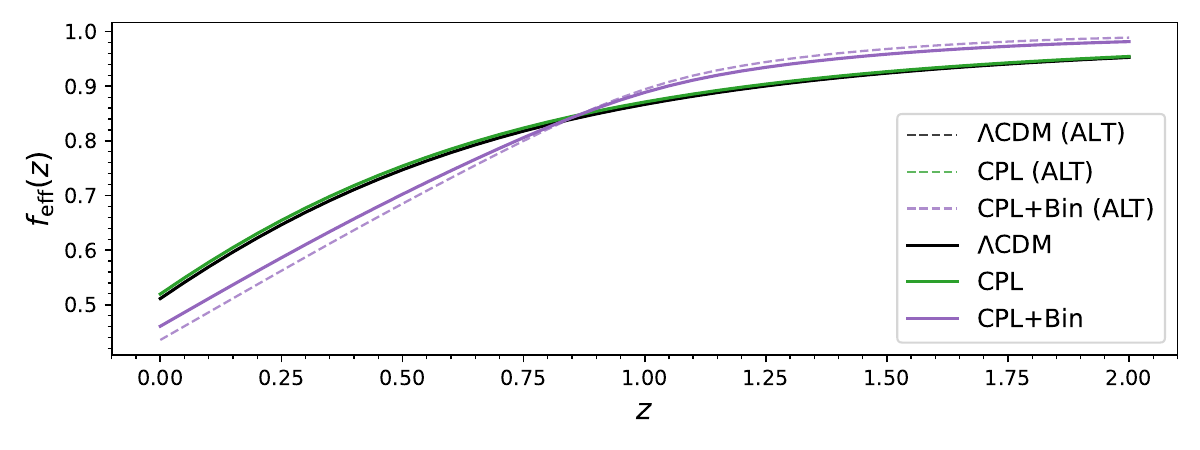}
    \includegraphics[scale=0.42]{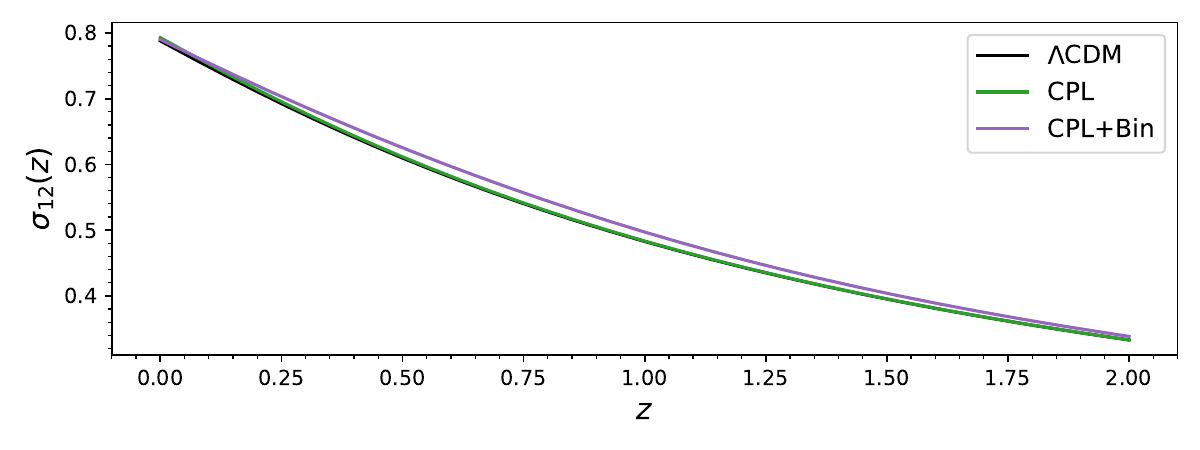}  
    \caption{{\it Upper plot:} Comparison of $f_{\rm eff}(z)$ \eqref{eq:feff} and $f_ {\rm eff,alt}(z)$ \eqref{eq:feff2} obtained with the best-fit $\Lambda$CDM, CPL and CPL+Bin models. For the former two these curves are identical; {\it Lower plot:} Comparison of $\sigma_{12}(z)$ for the same models. We use CAMB for the computation of these quantities.}\label{fig:f_eff}
\end{figure}

\begin{equation}\label{eq:feff}
    f_{\rm eff}(z)=\frac{\int_{0}^{\infty}dk\,k^2f(k,z)P_{\delta\delta}(k,z)W^2(kR_{12})}{\int_{0}^{\infty}dk\,k^2P_{\delta\delta}(k,z)W^2(kR_{12})}\,,
\end{equation}
which is the weighted average of the $k$-dependent growth rate

\begin{equation}
    f(k,z) \equiv \frac{d\ln \delta_m(k,z)}{d\ln a}
    = -\frac{(1+z)}{2} \frac{d\ln P_{\delta\delta}(k,z)}{dz}
\end{equation}
over the range of scales selected by the kernel function,

\begin{equation}\label{eq:kernel}
    K(k,z) = \frac{k^2P_{\delta\delta}(k,z)W^2(kR_{12})}{\int_{0}^{\infty}dk\,k^2P_{\delta\delta}(k,z)W^2(kR_{12})}\,,
\end{equation}
with 
\begin{equation}
    W(kR_{12})=\frac{3}{k^2R_{12}^2}\left[\frac{\sin(kR_{12})}{kR_{12}}-\cos(kR_{12})\right]\,.
\end{equation}

In Fig. \ref{fig:kernel}, one can see that the kernel rapidly decreases to negligible values at length scales smaller than those probed by galaxy surveys. Conversely, the contribution from scales larger than those measured by galaxy surveys to the effective growth rate \eqref{eq:feff} is also negligible because they only include a very limited range of small wave modes. Therefore, we conclude that the theoretical expression used in our main analyses already captures the relevant range of scales, namely those probed by RSD observations.

Nevertheless, one could also consider alternative physically motivated kernels. Ideally, the kernel should be tailored to the observational specifications of each galaxy survey, but considering the typical range of scales sensitive to RSD, we could simply replace the window function $W(kR_{12})$ -- which is obtained from a flat (top-hat) window function in real space -- with a top-hat window function in momentum space, e.g., in the wave-mode range $k\in[0.03,0.3]$ Mpc$^{-1}$, which corresponds to length scales between 21 and 210 Mpc. This would lead to the following alternative definition of the effective growth rate,

\begin{equation}\label{eq:feff2}
       f_{\rm eff, alt}(z)=\frac{\int_{0.03}^{0.3}dk \,k^2f(k,z)P_{\delta\delta}(k,z)}{\int_{0.03}^{0.3}dk\,k^2P_{\delta\delta}(k,z)}\,,
   \end{equation}
 which makes use of the kernel function  \begin{equation}\label{eq:kernel2}
K_{\rm alt}(k,z) = \frac{k^2P_{\delta\delta}(k,z)}{\int_{0.03}^{0.3}dk\,k^2P_{\delta\delta}(k,z)}\theta(k-0.03)\theta(0.3-k) \,,
\end{equation}
where $\theta(x)$ denotes the Heaviside step function ($=1$ for $x>0$ and $=0$ otherwise). In the upper plot of Fig. \ref{fig:kernel}, we compare the shape of this kernel to that of the kernel introduced in Eq. \eqref{eq:kernel}, and in Fig. \ref{fig:f_eff} we compare the curves of the effective growth rate for the various models and for the two definitions discussed in this appendix, as well as the curves of  $\sigma_{12}(z)$. We note that the difference between $\Lambda$CDM and CPL+Bin is slightly larger for $f_{\rm alt}(z)$ than for the CAMB/Planck-based definition of $f_{\rm eff}(z)$. This behavior is consistent with the kernel comparison shown in Fig.~\ref{fig:kernel}. The alternative kernel gives slightly more weight to the larger-$k$ side of the integral range. This is the region where the scale-dependent modification (suppression) in $\mu(k,z)$ is relevant, and therefore the deviation of the CPL+Bin model from $\Lambda$CDM becomes greater. In this sense, we find $f_{\rm eff}(z)$ to be more conservative, as it assigns less weight to mildly non-linear scales, which are less well under control. Nevertheless, the two kernels exhibit very similar behaviors, and consequently yield very similar values of the effective growth rate. We also note that both definitions exactly recover the true growth rate in the limit where it is independent of $k$, as required. This situation occurs, for example, in $\Lambda$CDM and CPL models with standard gravity, but it no longer applies when modified gravity effects are introduced. In such cases, the growth rate generally becomes scale-dependent, inheriting the $k$-dependence of the effective gravitational strength. See, e.g, our parametrized form of $\mu(k,z)$ in Eq. \eqref{eq:mu}.

Finally, we would like to emphasize that the suppression in the RSD observable $f\sigma_{12}(z)$ found in the CPL+Bin model does not originate from a decrease in the amplitude of the power spectrum, $\sigma_{12}(z)$, but rather from a decrease in the effective growth rate, as is again clear from the two panels of Fig.~\ref{fig:f_eff}.

\section{Significance of the new-physics signal}
In this appendix, we summarize the best-fit values of the cosmological parameters for all models considered in this work. The corresponding minimized chi-squared values for each dataset combination are also listed in Table \ref{table-best}. Besides, we additionally report the significance relative to the $\Lambda$CDM model, denoted by $E_{\Lambda{\rm CDM}}$ -- we compute it making use of the likelihood-ratio test, as done, e.g., in \cite{DESI:2025zgx,DESI:2025fii,Gonzalez-Fuentes:2025lei} --, together with the Akaike Information Criterion  difference, $\Delta {\rm AIC}\equiv$AIC$_{\Lambda{\rm CDM}}$-AIC$_{i}$ \cite{Akaike}. When deriving the best-fit points, we focus exclusively on the observational contributions from CMB, BAO, Type Ia supernovae and RSD, and ignore the priors associated with the Planck likelihood nuisance parameters. Therefore, the quoted $\chi^2_{\rm total}$ values correspond only to the observational likelihoods included in the analysis. 

Several notable trends that reinforce the discussion in the main text can be identified from Table~\ref{table-best}. First, introducing scale-dependent modified gravity effects significantly improves the fit to the RSD data, reducing $\chi^2_{\rm RSD}$ compared to the $\Lambda$CDM. This improvement is particularly pronounced for shorter critical wavelengths $\lambda_c$. 
Second, although the scale-dependent modified gravity models generally provide a better fit to large-scale structure observables, the improvement in the total $\chi^2$ is partially compensated by slightly worse fits to the CMB likelihood. 
Finally, allowing for a dynamical dark energy background significantly improves the fit to the DESI BAO and SNIa data. Overall, modified gravity in a dynamical dark energy background leads to the most favorable scenario.

As a result, for example, in the CPL+Bin (varying $\lambda_c$) case, we find a substantial improvement in the total fit, with $\Delta \chi^2_{\mathrm{total}} = 15.45$ and $\Delta {\rm AIC} =+ 5.45$, and with the likelihood ratio test pointing to an enhanced exclusion level of the $\Lambda$CDM, which goes from $2.31\sigma$ in the absence of MG effects to the $2.63 \sigma$ when the latter are taken into account. The signal raises to $2.80\sigma$ CL and $\Delta$AIC=+6.72 when we replace the SNIa from \texttt{Pantheon+} with those from DES-Dovekie -- see appendix C for further details.

The best-fit value, $\lambda_c \simeq 40\,{\rm Mpc}$, lies in the mildly non-linear regime and should therefore not be overinterpreted. In our view, the physically relevant result is not the precise best-fit value itself, but rather the preference for modifications entering on linear scales of order $50-100\,{\rm Mpc}$. Indeed, we find that the case $\lambda_c=62.8\,{\rm Mpc}$, already in the linear regime, also provides a comparably good fit to the data.
We therefore conclude that, if current cosmological datasets, including RSD measurements, are taken at face value, they favor modified gravity emerging on the linear scales proven by galaxy surveys. It is also worth noting that, given the the best-fit $\lambda_c\simeq 40\,\mathrm{Mpc}$, a more detailed treatment of non-linearities, for instance within the framework of the effective field theory of large-scale structure \cite{Carrasco:2012cv}, may become increasingly important in future studies.

\begin{figure}[t!]
    \centering
    \includegraphics[scale=0.44]{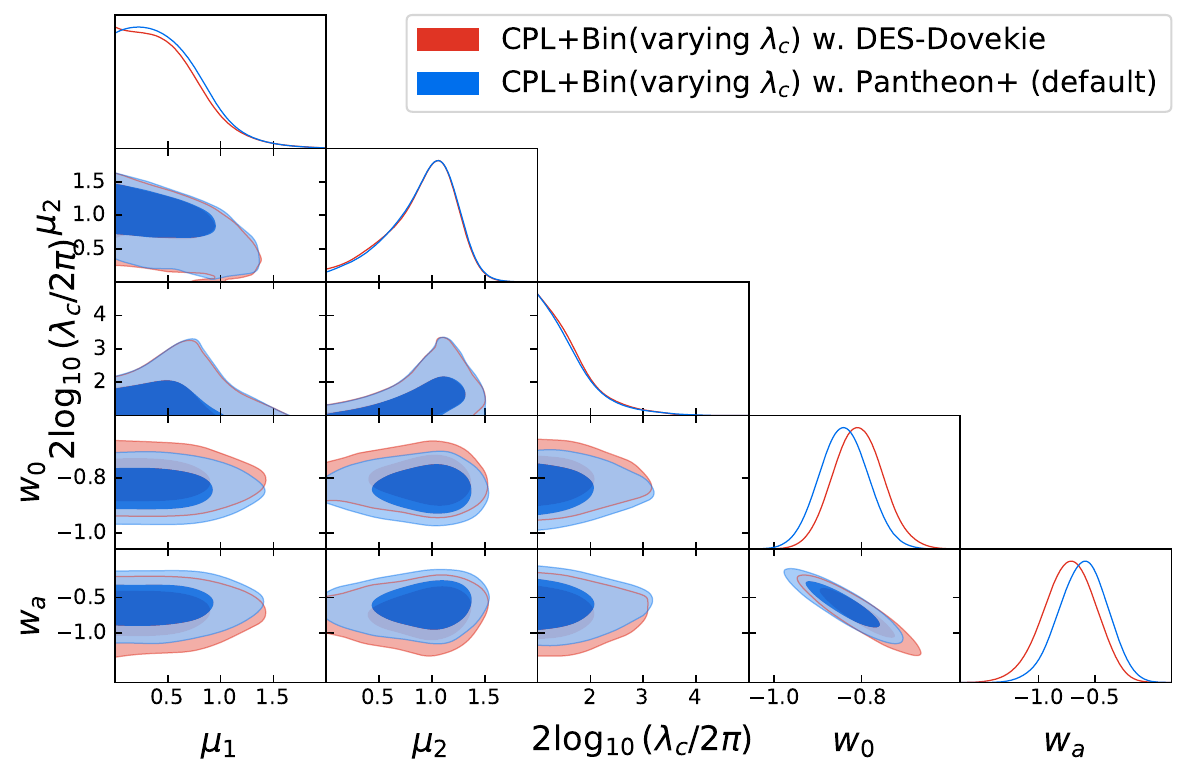}
    \caption{Triangle plots (at $68\%$ and $95\%$ CL) for the CPL background including MG effects, with varying $\lambda_c$. The red contours correspond to the analysis using the SNIa sample from DES-Dovekie, instead of Pantheon+. }
    \label{fig:Sd}
\end{figure}

\

\section{Comments on the results with DES-Dovekie}
As a consistency check, we also perform the analysis using DES-Dovekie~\cite{DES:2025sig} instead of \texttt{Pantheon+} as the Type Ia supernova catalog. We find that the overall constraints on the binning method parameters remain broadly consistent with those obtained using \texttt{Pantheon+}. For example, in the CPL+Bin scenario, we obtain $\mu_1 < 0.619$, $\mu_2 = 0.89^{+0.40}_{-0.22}$, and $\log_{10}\lambda_c < 1.66$ at 68\% CL. 
On the other hand, the differences between DES-Dovekie and \texttt{Pantheon+} mainly appear in the CPL background parameters. In the standard CPL case, we find $w_0 = -0.827 \pm 0.056$ and $w_a = -0.60^{+0.23}_{-0.20}$, while in the CPL+Bin case we obtain $w_0 = -0.808 \pm 0.057$ and $w_a = -0.72^{+0.24}_{-0.22}$. Compared to the analyses based on \texttt{Pantheon+}, DES-Dovekie enhances the signal of dynamical dark energy to $2.80\sigma$ CL, pushing the values of $w_0$ and $w_a$ slightly deeper into the phantom region. However, the overall tendency remains qualitatively similar, and the conclusions presented in the main body of this {\it Letter}, obtained with \texttt{Pantheon+}, remain unchanged.


\bibliographystyle{elsarticle-num}
\bibliography{ref}
\clearpage

\end{document}